\documentstyle[12pt]{article}

\newcommand{\bq}{\begin{equation}}
\newcommand{\eq}{\end{equation}}
\newcommand{\bqn}{\begin{eqnarray}}
\newcommand{\eqn}{\end{eqnarray}}
\newcommand{\nb}{\nonumber}
\newcommand{\lb}{\label} 

\begin{document} 
\title{Critical phenomena of collapsing massless scalar wave packets}
\author{ Anzhong Wang \thanks{e-mail address: wang@on.br; 
a.wang@vmesa.uerj.br}\\
\small Departamento de Astrof\'{\i}sica, Observat\'orio
Nacional~--~CNPq, \\ 
\small Rua General Jos\'e Cristino 77, S\~ao Crist\'ov\~ao,
20921-400 Rio de Janeiro~--~RJ, Brazil\\
  \\
Henrique P. de Oliveira \\
\small Departamento de F\' {\i}sica Te\' orica,
Universidade do Estado do Rio de Janeiro, \\
\small Rua S\~ ao Francisco Xavier 524, Maracan\~ a,
20550-013 Rio de Janeiro~--~RJ, Brazil}

\maketitle

\begin{abstract}

\baselineskip 0.6cm

An analytical model that represents the collapse of a massless scalar
wave packet with continuous self-similarity is constructed, and
critical phenomena are found. In the supercritical case, the mass of
black holes is finite and has the form $M \propto (p - p^{*})^{\gamma}$,
with $\gamma = 1/2$.
 
\end{abstract}

\vspace{.4cm}


\newpage

\baselineskip 0.9cm

\section*{I. Introduction}

Recently, Choptuik \cite{Ch1993} studied the collapse of a massless
scalar field with spherical symmetry, and found the following
intriguing features:  Let the initial distribution of the massless
scalar field be parametrized smoothly by a parameter $p$ that
characterizes the strength of the initial conditions, such that the
collapse of the scalar field with the initial data $p > p^{*}$  forms a
black hole, while the one with  $p < p^{*}$ does not. Then, it was
found that:  (i) the critical solution with $p = p^{*}$ is {\em
universal} in the sense that in all the one-parameter families of the
solutions considered it approaches an identical spacetime; (ii) the
critical solution has a discrete self-similarity (DSS); (iii) near the
critical solution (but with $p > p^{*}$), the black hole mass is given
by 
$$ M_{BH} = K (p - p^{*})^{\gamma}, 
$$ 
where $K$ is a family-dependent constant, but $\gamma$ is an apparently
{\em universal} scaling exponent, which has been numerically determined
as $\gamma \approx 0.37$.  These phenomena were soon  found also in the
collapse of axisymmetric gravitational waves \cite{AE1993} as well as
in the one of radiation fluid \cite{EC1994}.  Therefore, it seems that
the phenomena are not due to the particular choice of the matter
fields but rather are generic features of General Relativity.
Further numerical evidences to support this conclusion are given in
\cite{EGH1995}.

Parallelly to the above numerical investigations, there have been
analytical efforts to understand the physics behind these phenomena
\cite{Brady1994,ONT1994,HMN1994,Maison1995, Gundlach1995,OC1996}. While
the universality of the critical solution and its self-similarity
(echoing) have been found in most cases considered, there exists
dispute over the universal scaling exponent $\gamma$.  Maison
\cite{Maison1995} showed that $\gamma$ is matter-dependent. For the
collapse of the perfect fluid with the equation of state $P = k \rho$,
it strongly depends on $k$, where $P$ and $\rho$ are respectively the
pressure and energy density of the fluid, and $k$ is a constant.  This
dependence is further shown in \cite{OC1996}. The same conclusion was
also obtained numerically by Eardley and co-workers \cite{EHH1995}.
Thus, one might expect that $\gamma$ is universal only within a
particular family of matter fields.  However, even in this sense the
analytical results are not consistent with the numerical ones.  In
particular, it was  shown \cite{Brady1994,ONT1994} that for the
massless scalar field the exponent $\gamma$ is $0.5$, instead of its
numerical value $\gamma \approx 0.37$.  Moreover, Maison
\cite{Maison1995} showed that the value $k = 0.88$ is the maximal one
for which a regular self-similar solution could be found for the
perfect fluid.  As we know, a massless scalar field is energetically
equivalent to a perfect fluid with $k = 1$ \cite{TW1991}, thus, Maison's
results seem also in conflict with the ones obtained in
\cite{Brady1994,ONT1994}.

It should be noted that the model considered in
\cite{Brady1994,ONT1994}, which will be referred to as the BONT model,
is not asymptotically flat, and the exponent $\gamma$ is obtained in a
rather unusual way. Moreover, except for the one of Gundlach
\cite{Gundlach1995}, all the analytic models, including the BONT model,
have continuous self-similarity (CSS), in contrast to the original
 model of Choptuik that has DSS \cite{Ch1993}. Thus, it is not clear
whether the difference in $\gamma$ obtained above is due to the
different self-similarities or due to the non-asymptotic flatness of
the spacetime. If it is because of the former, we will have a new
classification to the critical behavior, and many new solutions that
represent critical phenomena with different similarities  need to be
found. To resolve the above problem, in this paper we shall make a
``surgery" to the BONT model \cite{ONT1994}. That is,
we shall cut the spacetime along a null hypersurface and then join it
to an out-going Vaidya solution, so the resulted spacetime is
asymptotically flat and has a finite  mass.  Clearly, such obtained
solution will represent the collapse of a massless scalar wave packet,
which is radiating as it collapses. This model is more realistic and
more comparable with the one of Choptuik but with different
self-similarities. Specifically, the paper is organized as follows:  In
Sec. II, we briefly review the main properties of the BONT model, while
in Sec. III we cut the BONT spacetime along a null hypersurface and
then join it with an out-going Vaidya solution. In Sec. IV we consider
a particular case of the solutions obtained in Sec. III, which
represents the collapse of a scalar wave packet. Critical phenomena are
found with the exponent $\gamma = 1/2$. Finally, the paper is closed by
Sec. V, where our main conclusions are presented.

\section*{II. The BONT model}

The BONT model is described by the solutions \footnote{This class of
solutions was first found in \cite{Roberts1989}, but their physical
interpretation of representing critical phenomena was first given in
\cite{Brady1994,ONT1994}. The Roberts solutions are continuous
self-similar, since they possess the Killing vector, $X = u\partial u +
v\partial v$ with the property $L_{X}g_{\mu\nu} = 2 g_{\mu\nu}$, where
$L_{X}$ denotes the Lie derivative with respect to $X$.}
\bq
\lb{eq1}
ds^{2} = - G(u, v) du dv + r^{2}(u, v) d^{2}\Omega, 
\eq
where $d^{2}\Omega \equiv  d\theta^{2} + \sin^{2}\theta d\varphi^{2}$,
and  the metric coefficients are given by
\bqn
\lb{eq2}
r(u, v) &=& \frac{1}{2}\left[ u^{2}  - 2uv + 4 b_{2}v^{2}
\right]^{1/2} H(v) + \frac{1}{2}\left[2(a(v) - a(0)) -  
u \right] H(- v), \nb\\
G(u, v) &=& H(v) + 2 a'(v) H(- v),
\eqn
where $b_{2}$ is an arbitrary constant, and $H(x)$ denotes the
Heaviside function, which is one for $x \ge 0$ and zero for  $x < 0$. A
prime denotes the ordinary differentiation, and $a(v)$ is an arbitrary
function subject to $a'(v) > 0$ and $a'(0) = 1/2$. One can show that
the hypersurface $v = 0$ is free of any matter and represents a
boundary surface \cite{Israel1966}. The corresponding massless scalar
field is given by
\bq
\lb{eq3}
\phi = \pm \frac{1}{\sqrt{2}}\ln\left|\frac{(u - v) - \sqrt{1 - 4b_{2}}v}
{(u - v) + \sqrt{1 - 4b_{2}}v}\right| H(v).
\eq
Note the slight difference in the notations used here and the ones used
in \cite{Brady1994,ONT1994}. From the above expressions one can show
that the spacetime is Minkowski in the region $u < 0, v < 0$, while in
the region $u < 0, v > 0$ it represents a collapsing massless scalar
wave. When $b_{2} < 0$, the scalar wave collapses into a spacetime
singularity on the hypersurface $ u = - [\sqrt{1 - 4b_{2}} - 1]v$,
which is preceded by  an apparent horizon at $u = 4b_{2}v$.  Thus, the
corresponding solutions represent the formation of black holes and are
supercritical. When $b_{2} = 0$, the singularity coincides with the
apparent horizon  on $u = 0$ and becomes null. This  
solution is critical, which separates the supercritical solutions from
the subcritical ones. The subcritical solutions are these with  $0 <
b_{2} < 1/4$. It can be shown that in the latter case the scalar wave
first collapses and then disperses into infinity, without forming black
holes but leaving behind a Minkowski region $u, v > 0$, in which the
metric takes the form of Eq.(\ref{eq1}) with
\bqn
\lb{eq4}
G(u, v) &=& 4\sqrt{b_{2}}\; b'(u) , \;\;\;\; r = \sqrt{b_{2}}\;v 
- \left[b(u) - b(0)\right], \nb\\
\phi(u, v) &=& \pm \frac{1}{\sqrt{2}}\ln
\left[\frac{1 + \sqrt{1 - 4b_{2}}}{1 - \sqrt{1 - 4b_{2}}}\right], 
\;\; (u, v > 0)  
\eqn
where $b(u)$ is an arbitrary function, subject to $b'(u) > 0, b'(0) =
1/(4\sqrt{b_{2}})$. One can show that the hypersurface $u = 0, v >
0$ is also a boundary surface. 

In the region $u <0, v > 0$, the local mass of the scalar field
is given by
\bq
\lb{eq5}
m(u, v) = \frac{r}{2}\left(1 - r_{,\alpha}r_{,\beta} 
g^{\alpha \beta}\right) = - \frac{(1 - 4b_{2})uv}{8r}, 
\eq
where $r$ is given by Eq.(\ref{eq2}). Clearly, on the apparent horizon
$u = 4b_{2}v$ the mass becomes unbounded as $v \rightarrow + \infty$
for $b_{2} < 0$. That is, the spacetime fails to be asymptotically flat
for the supercritical case. As a result, the total mass of the black
hole can not be written in a power-law form in terms of initial data,
whereby the exponent $\gamma$ can be read out. For more details, we
refer the readers to \cite{Brady1994,ONT1994}.

\section*{III. Matching the Roberts solutions to the out-going Vaidya
dust solution}

To circle the above problem, we shall cut the spacetime along the
hypersurface, say, $v = v_{0} > 0$, and keep the region  $v \le v_{0}$,
while the region $v \ge v_{0}$ will be replaced by that of out-going
dust Vaidya solution \cite{Vaidya1951}. To do the matching, following
Barrab\'es and Israel \cite{BI1991}, we first write the Roberts
solutions in the form 
\bq
\lb{eq6}
ds^{2}_{-} = - e^{\psi_{-}}dv (f_{-} e^{\psi_{-}}dv - 2dr) 
+ r^{2}d^{2}\Omega, 
\eq
where $v$ is the Eddington advanced time, and for Roberts' solutions 
(\ref{eq2})
in the region $u < 0, v > 0$, the metric coefficients are given by
\bqn
\lb{eq7}
f_{-} &=& \frac{\left[(1 - 4b_{2})v^{2} + 4r^{2}\right]^{1/2}}{4r^{2}}
\left\{\left[(1 - 4b_{2})v^{2} + 4r^{2}\right]^{1/2} -
(1 - 4b_{2})v\right\}, \nb\\
e^{\psi_{-}} &=& \frac{2r}{\left[(1 - 4b_{2})v^{2} + 
4r^{2}\right]^{1/2}}.
\eqn
Now we restrict Eq.(\ref{eq7}) valid only in the region $u < 0, 0 \le v
\le v_{0}$. Then, the normal to the hypersurface $\Phi^{-} = v - v_{0}
= 0$ is $n^{-}_{\mu} = \alpha^{- 1}\partial_{\mu}\Phi^{-} = \alpha^{-
1}\delta^{v}_{\mu}$, where $\alpha$ is a negative function. From
$n^{-}_{\mu}$ we can introduce a ``transverse" null vector
$N^{-}_{\mu}$ by requiring $N^{-}_{\lambda}N^{- \lambda} = 0$, and
$N^{-}_{\lambda}n^{- \lambda} = - 1$. Without loss of generality, we
assume that $N^{-}_{\mu}$ takes the form $N^{-}_{\mu} =
N^{-}_{v}\delta^{v}_{\mu} + N^{-}_{r}\delta^{r}_{\mu}$, and choose the
arbitrary function $\alpha$ as $\alpha = - exp\{- \psi_{-}\}$. Then, it
is easy to show that $N^{-}_{\mu}$ is given by
\bq
\lb{eq8}
N^{-}_{\mu} = - \frac{f_{-}}{2}e^{\psi_{-}}\delta^{v}_{\mu} + 
\delta^{r}_{\mu}.
\eq

Choosing the coordinates $r, \theta$, and $\varphi$ as the three
intrinsic coordinates $\xi^{a} \equiv (r, \theta, \varphi), \; (a = 1,
2, 3)$ on the hypersurface $v = v_{0}$, we find
\bq
\lb{eq9}
e^{- \mu}_{(1)} = \delta^{\mu}_{r},\;\;\;
e^{- \mu}_{(2)} = \delta^{\mu}_{\theta},\;\;\; 
e^{- \mu}_{(3)} = \delta^{\mu}_{\varphi},
\eq
where $e^{- \mu}_{(a)} \equiv \partial x^{\mu}_{-}/\partial\xi^{a}$.
Then, it can be shown that the ``transverse" extrinsic curvature,
defined by \cite{BI1991}
\bq
\lb{eq10}
{\cal{R}}_{a b} = - N_{\mu} e^{\nu}_{(b)} 
\left(\nabla_{\nu}e^{\mu}_{(a)}\right),
\eq
takes the form
\bq
\lb{eq11}
{\cal{R}}^{-}_{a b} = diag.\left\{ - \frac{\partial \psi_{-}}{\partial r},
\; \frac{r f_{-}}{ 2},\; \frac{r f_{-}}{2}\sin^{2}\theta\right\}.
\eq
Note that in calculating the above equation, we have not used the
particular expressions (\ref{eq7}) for the functions $\psi_{-}$ and
$f_{-}$.  Thus, it is valid for the general case.

On the other hand, the out-going Vaidya solution \cite{Vaidya1951} can
be written in the form
\bq
\lb{eq12}
ds^{2}_{+} = - e^{\psi_{+}}dU (f_{+} e^{\psi_{+}}dU + 2dr) 
+ r^{2} d^{2}\Omega, 
\eq
where
\bq
\lb{eq13}
f_{+} = 1 - \frac{2 m(U)}{r},\;\;\; \psi_{+} = 0,
\eq
and $U$ is the Eddington retarded time, which is in general the
function of $u$ appearing in Eq.(1), and $m(U)$ is the local mass of
the out-going Vaidya dust.  The corresponding energy-momentum tensor is
given by
\bq
\lb{eq14}
T^{+}_{\mu\nu} = - \frac{2}{r^{2}}\frac{dm(U)}{dU} \delta^{U}_{\mu}
\delta^{U}_{\nu}.
\eq
In the following, we shall take metric (\ref{eq12}) as valid only in
the region $u < 0, v \ge v_{0}$. To have our results more applicable,
for the moment we shall not restrict ourselves to the particular
solution (\ref{eq13}). The hypersurface $v = v_{0}$ in the ($U,
r$)-coordinates can be written as $\Phi^{+} = U - U_{0}(r) = 0$, where
$U_{0}(r)$ is a solution of the equation
\bq
\lb{eq16}
\frac{dU_{0}}{dr} =  - \frac{2}{f_{+}} e^{- \psi_{+}}, \;\;\; (v = v_{0}).
\eq
Then, the normal to the surface is given by 
$$
n^{+}_{\mu} = \beta^{- 1}\partial_{\mu}\Phi^{+} = 
\beta^{- 1}\left(\delta^{U}_{\mu} + \frac{2}{f_{+}}e^{ - 
\psi_{+}}\delta^{r}_{\mu}\right),
$$
where $\beta$ is a negative otherwise arbitrary function. From
$n^{+}_{\mu}$ we can also introduce the ``transverse" null vector
$N^{+}_{\mu}$, by requiring $N^{+}_{\lambda}N^{+ \lambda} = 0$, and
$N^{+}_{\lambda}n^{+ \lambda} = - 1$. It can be shown that it takes the
form
\bq
\lb{eq17}
N^{+}_{\mu} = \frac{\beta f_{+}}{2} e^{2\psi_{+}}\delta^{U}_{\mu}.
\eq
On the other hand, we also have
\bq
\lb{eq18}
e^{+ \mu}_{(1)} = - \frac{2}{f_{+}}e^{- \psi_{+}}\delta^{\mu}_{U} 
+ \delta^{\mu}_{r},\;\;\;
e^{+ \mu}_{(2)} = \delta^{\mu}_{\theta},\;\;\; 
e^{+ \mu}_{(3)} = \delta^{\mu}_{\varphi},
\eq
where $e^{+ \mu}_{(a)} \equiv \partial x^{\mu}_{+}/\partial\xi^{a}$. To
be sure that the two ``transverse" vectors $N^{\pm}_{\mu}$ defined in
the two faces of the hypersurface $v = v_{0}$ represent the same
vector, we need to impose the condition
$$
N^{+}_{\lambda} e^{+ \lambda}_{(a)}\left|_{v = v_{0}} = 
N^{-}_{\lambda} e^{- \lambda}_{(a)}\right|_{v = v_{0}},
$$
which requires that the function $\beta$ has to be $\beta = - exp\{-
\psi_{+}\}$. Once $N^{+}_{\lambda}$ and $e^{+ \lambda}_{(a)}$ are
given, using Eq.(\ref{eq10}) we can calculate the corresponding
``transverse" extrinsic curvature, which in the present case takes the
form
\bq
\lb{eq19}
{\cal{R}}^{+}_{a b} = diag.\left\{ - 
\frac{2e^{- \psi_{+}}}{f_{+}^{2}}\frac{\partial f_{+}}{\partial U},\; 
\frac{r f_{+}}{2},\; \frac{r f_{+}}{2} \sin^{2}\theta\right\}.
\eq
Then, from Eqs.(\ref{eq11}) and (\ref{eq19}), we find that
\bqn
\lb{eq20}
\gamma_{a b} &=& 2 \left({\cal{R}}^{+}_{a b} - 
{\cal{R}}^{-}_{a b}\right) \nb\\
&=& \left[\frac{2\partial \psi_{-}}{\partial r} - 
\frac{4 e^{- \psi_{+}}}{f^{2}_{+}}\frac{\partial f_{+}}{\partial U}
\right]
\delta^{r}_{a}\delta^{r}_{b}
+ r(f_{+} - f_{-})(\delta^{\theta}_{a}\delta^{\theta}_{b} + 
\sin^{2}\theta \delta^{\varphi}_{a}\delta^{\varphi}_{b}).
\eqn
Once $\gamma_{a b}$ is given, using the formula \cite{BI1991}
\bq
\lb{eq21}
\tau^{a b} = - S^{a b} = \frac{1}{16 \pi}(g^{a c}_{*} l^{b}l^{d}
+ g^{b d}_{*} l^{a}l^{c} - g^{a b}_{*} l^{c}l^{d}
- g^{c d}_{*} l^{a}l^{b})\gamma_{c d},
\eq
we can calculate the surface energy-momentum tensor $\tau^{a b}$ on the
null hypersurface $v = v_{0}$, which now can be
written as
\bq
\lb{eq22}
\tau^{a b} = \sigma l^{a}l^{b} + P g^{a b}_{*},
\eq
where
\bqn
\lb{eq23}
\sigma &=& - \frac{f_{+} - f_{-}}{8\pi r},\nb\\
P &=&  \frac{1}{8\pi}\left(\frac{2e^{- 
\psi_{+}}}{f_{+}^{2}} \frac{\partial f_{+}}{\partial U}
- \frac{\partial\psi_{-}}{\partial r}\right),
\eqn
and 
\bqn
\lb{eq24}
g^{a b}_{*} &=& r^{- 2}\left(\delta^{a}_{\theta}\delta^{b}_{\theta}
+ \sin^{- 2}\theta\delta^{a}_{\varphi }\delta^{b}_{\varphi}\right),\nb \\
l^{a} &=& \delta^{a}_{r},\;\;\;\;\;\; l^{b}l_{b} = 0.
\eqn
The function $\sigma$ represents the surface energy density of the null
shell, and $P$ the pressures in the $\theta$- and
$\varphi$-directions.  Note that in \cite{BI1991} the case of a null
shell was also considered. But, there they used the same null
coordinate in both sides of the shell, while here  we use the retarded
null coordinate in one side of the shell and the advanced null
coordinate in the other side  [cf. Eqs.(\ref{eq6}) and (\ref{eq12})].

For the particular solutions given by Eqs.(\ref{eq7}) and (\ref{eq13}),
Eq.(\ref{eq23}) yields  
\bqn
\lb{eq25}
\sigma &=& \frac{1}{4\pi r^{2}}\left\{M(r) - \frac{(1 - 4b_{2})v_{0}}{8 r}
\left[\sqrt{(1 - 4b_{2})v_{0}^{2} + 4 r^{2}} - v_{0}\right]\right\}, \nb\\
P &=& \frac{1}{4\pi (r - 2M)}\left\{\frac{dM(r)}{dr}
 - \frac{(1 - 4b_{2})v_{0}^{2}(r - 2M)}
{2r[(1 - 4b_{2})v_{0}^{2} + 4 r^{2}]}\right\},
\eqn
where
\bq
\lb{eq26}
M(r) \equiv \left. m(U)\right|_{U = U_{0}(r)},
\eq
and $U_{0}(r)$ is a solution of Eq.(\ref{eq16}).

\section*{IV. Critical phenomena for the case P = 0}

To study the general shell given by Eq.(\ref{eq25}), it is found very
complicate. In this section, we shall consider the case where $P = 0$,
i.e.,
\bq
\lb{eq27}
\frac{dM(r)}{dr} =  \frac{(1 - 4b_{2})v_{0}^{2}(r - 2M)}
{2r[(1 - 4b_{2})v_{0}^{2} + 4 r^{2}]}.
\eq
Integrating the above equation, we find that
\bq
\lb{eq28}
 M(r)  =  \frac{1}{r}\left\{ p\sqrt{4p^{*^{2}} + r^{2}} - 
2 p^{*^{2}}\right\},  
\eq
where $p$ is the integration constant, and
\bq
\lb{eq29}
p^{*} \equiv \frac{\sqrt{1 - 4b_{2}}}{4}v_{0}.
\eq
At the past null infinity ($r \rightarrow + \infty$), Eq.(\ref{eq28})
shows that
\bq
\lb{eq30}
M(r \rightarrow + \infty) = p. 
\eq
That is, the parameter $p$ represents the total initial mass of the
massless scalar wave packet and the null shell, with which they
collapse. As $r \rightarrow 0^{+}$, $M(r)$ has the following asymptotic
behavior
\bq
\lb{eq31}
M(r) \rightarrow \left\{ \begin{array}{ll}
+ \infty, &  p > p^{*}, \\
0, &  p = p^{*}, \\
- \infty, &  p < p^{*}. \end{array} \right.
\eq
On the other hand, it is well-known that the apparent horizon at $r =
2M(r)$ of the out-going Vaidya solution always coincides with its
future event horizon. Thus, by comparing the mass $M(r)$ with
$r/2$ we can tell whether the collapse forms a black hole or
not,
\bq
\lb{eq32}
 M(r)  - \frac{r}{2} =  \left\{\frac{4p^{*^{2}} + r^{2}}
 {4r^{2}[p + (4p^{*^{2}} + r^{2})]}\right\}^{1/2}
\left[4(p^{2} - p^{*^{2}}) - r^{2}\right].
\eq
Clearly, only when $p > p^{*}$, the scalar field and the null shell
will collapse inside the event horizon at
\bq
\lb{eq33}
r_{AH} = 2 \sqrt{p^{2} - p^{*^{2}}}.   
\eq
When $p = p^{*}$, $M(r) = r/2$ is possible only at the origin,
$r = 0$, where a zero-mass singularity is formed. Thus, the solution
with $p = p^{*}$ represents the critical solution that separates the
supercritical solutions ($p > p^{*}$) from the subcritical ones ($p <
p^{*}$). In the subcritical case, $M(r)$ is always greater than
$r/2$, and the collapse never forms a black hole [cf. Fig.1].
In the latter case, the region $u, v > 0$ should be replaced by the
Minkowski solution (\ref{eq4}).  As shown in the last section, the
matching across the hypersurface $u = 0, \; 0 \le v \le v_{0}$ is
smoothy, i.e., no matter appears on it. To show that it is also the case
on the hypersurface $u = 0, \; v \ge v_{0}$, which separates the Vaidya
solution (\ref{eq13}) from the Minkowski (\ref{eq4}), we first make the
coordinate transformation $U = U(u)$, and then write the metric
(\ref{eq12}) in terms of $u$. Using the results obtained in
\cite{BI1991}, one can show that to have a smooth matching we have to
impose the condition
$$
U'(0) = \frac{1}{\sqrt{4b_{2}}},\;\;\; M(r)\left|_{u = 0} = 0\right..
$$
Clearly, by properly choosing the function-dependence of $U$ on $u$,
the first condition can be always satisfied. On the other hand, from
Eqs.(\ref{eq2}) and (\ref{eq28}) one can show that the last condition
is also satisfied identically. Therefore, the matching of the Vaidya
solution to the Minkowski one across the hypersurface $u = 0, \; v \ge
v_{0}$ is always possible for $p < p^{*}$. The corresponding Penrose
diagram for each of the three cases are shown in Fig. 2.

On the apparent horizon $r = r_{AH}$, the total mass of the scalar wave
packet and the shell is
\bq
\lb{eq34}
M_{BH} = \frac{r_{AH}}{2} = K (p - p^{*})^{1/2},  
\eq
where $K \equiv (p + p^{*})^{1/2}$. The above expression shows that
the black hole mass takes a power-law form with its exponent $\gamma$
being equal to 0.5. This is exactly the same value as that obtained in
\cite{Brady1994,ONT1994}. This result seems a little bit surprising,
since our model is different from the one of \cite{Brady1994,ONT1994}.
In particular, in our case the collapse consists of two parts, one is
the massless scalar wave packet and the other is the null shell, while
in \cite{Brady1994,ONT1994} only a collapsing scalar wave exists.
However, a more detailed investigation shows that the existence of the
shell does not affect our main conclusions. In fact, by properly
choosing the parameters in our model the shell can be made completely
disappear, without affecting the main properties of the collapse.  To
show this, let us write the total mass $M(r)$ as
\bq
\lb{eq35}
M(r) = M_{\phi} + M_{shell},
\eq
where $M_{\phi}$ denotes the total mass of the scalar wave packet, and 
$M_{shell}$ the total mass of the shell, given respectively by
\bqn
\lb{eq36}
 M_{\phi} &=& \frac{2p^{*^{2}}}{v_{0} r}
\left\{2 \sqrt{4 p^{*^{2}} + r^{2}} - v_{0}\right\},
\nb \\
M_{shell} &=&\frac{v_{0}p - 4p^{*^{2}}}{v_{0}}
\left\{\frac{4 p^{*^{2}} + r^{2}} {r^{2}}\right\}^{1/2}.
\eqn
Clearly, if we choose the parameters such that
\bq
\lb{eq37}
v_{0} = \frac{4 p^{*^{2}}}{p},  
\eq
the null shell disappears. Moreover, using Eqs.(\ref{eq29}) and
(\ref{eq37}), we find
\bq
\lb{eq38}
p - p^{ *}   = \left[\frac{4 p^{*^{2}}}{p + p^{*}}\right](- b_{2}).  
\eq
Therefore, when the null shell is absent, the massless scalar wave
packet will collapse to form a black hole for $b_{2} < 0$, and first
collapses and then disperses to infinity, without forming a black hole
for $1/4 > b_{2} > 0$ \cite{Brady1994,ONT1994}.

\section*{V. Concluding remarks}

In this paper, by ``cut-paste" method, we have constructed a physically
more realistic model for the collapse of a massless scalar wave packet,
which is usually accompanied by a collapsing null dust shell. Critical
phenomena are also found in these solutions, and the mass of the black
holes is finite and takes the form $M \propto (p - p^{*})^{\gamma}$.
In our model, the exponent $\gamma$ is 0.5, which is the same as that
found in \cite{Brady1994,ONT1994}, but different from the numerical one
of Choptuik, which is  0.37. However, the results do not contradict and
rather show the fact that $\gamma$ is not only matter-dependent but
also symmetry-dependent. In Choptuik's model, the critical solution has
DSS, while in \cite{Brady1994,ONT1994} and ours it has CSS.

On the other hand, quite recently Choptuik, Chmaj, and Bizo\'n
\cite{CCB1996} studied the collapse of a Yang-Mills field, and found
two distinct critical solutions at the threshold of black hole
formation. In one case, the critical solution has DSS, and the mass of
black holes for the supercritical solutions takes a power-law form
with the exponent $\gamma \approx 0.20$. This class of solutions was
referred to as Type II solutions. In the other case, the critical
solution is the static $n = 1$ Bartnik-Mckinnon sphaleron
\cite{BM1988}, and the formation of black holes always turns on at finite
mass. This class of solutions was referred to as Type I solutions. 

Summarizing all the above results, the following seems to emerge: The
solutions of gravitational collapse can be at least divided into three
different types, a) Type II.A solutions. In these solutions, the mass
of black holes always takes the form $M \propto (p -
p^{*})^{\gamma_{A}}$. b) Type II.B solutions. In this class of
solutions, the black hole mass also takes the power-law form, $M
\propto (p - p^{*})^{\gamma_{B}}$, but the only difference to Type II.A
is that the critical solution of Type II.A has DSS, while the one of
Type II.B has CSS. Because of this difference, the exponent $\gamma$
is  in general also different. c) Type I solutions. These solutions have
no self-similarities, neither DSS nor CSS, and the formation of black
holes always turns on at finite mass. If the above classification is
universal, a natural question is that:  Do these three types of
solutions saturate all the solutions that represent the formation of
black holes in gravitational collapse?


\section*{Acknowledgment}

The financial assistance from CNPq is gratefully acknowledged.


\newpage
\section*{Figure Captions}

Fig. 1   The qualitative behavior of the total mass
$M(r)$ defined by Eq.(\ref{eq28}) in the text for the three different
cases: (a) The supercritical case ($p > p^{*}$); (b) the critical case
($p = p^{*}$); and (c) the subcritical case ($p < p^{*}$).

Fig. 2  The Penrose diagram of the solutions with $P = 0$ for the three
different cases: (a) The supercritical case; (b) the critical case; and
(c) the subcritical case. Double lines represent spacetime
singularities.


\begin{thebibliography}{100}


\bibitem{Ch1993} M.W. Choptuik, Phys. Rev. Lett. {\bf 70}, 9 (1993).

\bibitem{AE1993} A.M. Abrahams and C.R. Evans, Phys. Rev. Lett. {\bf
70}, 2980 (1993); Phys. Rev. {\bf D49}, 3998 (1994).

\bibitem{EC1994} C.R. Evans and J.S. Coleman, Phys. Rev. Lett.  {\bf
72}, 1782 (1994).

\bibitem{EGH1995} E.W. Hirschmann and D.M. Eardley, Phys. Rev. {\bf
D51}, 4198 (1995); {\em ibid.} {\bf D52}, 5850 (1995); D. Garfinkle,
{\em ibid.}, {\bf D51}, 5558 (1995); R.S. Hamad\'e and J.M.
Stewart,Class. Quantum Grav.  {\bf 13}, 497 (1996).

\bibitem{Brady1994} P.R. Brady, Class. Quantum Grav. 11, 1255 (1994);
Phys. Rev. {\bf D51}, 4168 (1995).

\bibitem{ONT1994} Y. Oshiro, K. Nakamura, and A. Tomimatsu, Prog.
Theor. Phys. {\bf 91}, 1265 (1994).

\bibitem{HMN1994} V. Husain, E.A. Martinez, and D.  Nunez, Phys. Rev.
{\bf D50}, 3783 (1994); J. Trachen, {\em ibid.} {\bf D50}, 7144 (1994);
T. Koike and T. Mishima,  {\em ibid.} {\bf D51}, 4045 (1995); T. Koike,
T. Hara, and S. Adachi, Phys. Rev. Lett.  {\bf 74}, 5170 (1995);
J.F.  Da Rocha, N.O. Santos, and A.Z. Wang, ``{\em Spherical and
Conformally-Flat Spacetimes for Massless Scalar and Radiation Fields:
Analytical Solutions}," preprint (1996).

\bibitem{Maison1995} D. Maison, Phys. Lett. {\bf B366}, 82 (1996).

\bibitem{Gundlach1995} C. Gundlach,  Phys. Rev. Lett.  {\bf 75}, 3214
(1995); ``{\em Understanding Critical Collapse of a scalar Field},"
gr-qc/9604019.

\bibitem{OC1996} H.P. de Oliveira and E.S. Cheb-Terrab, Class. Quantum
Grav. {\bf 13}, 425 (1996); H.P. de Oliveira, ``{\em Self-Similar
Collapse in Brans-Dicke Theory and Critical Behavior},"
gr-qc$/$9605008; T. Chiba and J. Soda, ``{\em Critical Behavior in the
Brans-Dicke Theory of Gravitation}," gr-qc$/$9603056.

\bibitem{EHH1995}D.M. Eardley, E.W. Hirschmann, and J.H. Horne, Phys.
Rev. {\bf D52}, {\bf R}5397 (1995);  R.S. Hamade, J.H. Horne, and J.M.
Stewart, ``{\em Continuous self-similarity and S-Duality,}"
gr-qc/9511024; E.W. Hirschmann and D.M. Eardley,
 ``{\em Criticality and Bifurcation in the Gravitational Collapse of a
Self-Coupled Scalar Field},"   gr-qc/9511052.


\bibitem{TW1991} D. Tsoubelis and A.Z. Wang, J. Math. Phys. {\bf 32},
1017 (1991).

\bibitem{Roberts1989} M.D. Roberts, Gen. Rel. Grav. {\bf 21}, 907 (1989).

\bibitem{Israel1966} W. Israel, Nuovo Cimento, {\bf B44}, 1 (1966); {\em
ibid.}, {\bf B48}, 463(E) (1967).

\bibitem{Vaidya1951} P.C. Vaidya, Proc. Ind. Acad. Sciences, {\bf A33},
264 (1951).

\bibitem{BI1991} C. Barrab\'es and W. Israel, Phys. Rev. {\bf D43}, 1129
(1991).

\bibitem{CCB1996} M.W. Choptuik, T. Chmaj, and P. Bizo\'n, ``{\em Critical
Behavior in Gravitational Collapse of A Yang-Mills Field},"
gr-qc$/$9603051.
\bibitem{BM1988} R. Bartnik and J. Mckinnon, Phys. Rev. Lett. 
{\bf 61}, 141 (1988).

\end{thebibliography}
\end{document}